\documentclass[preprintnumbers, prd, onecolumn, floatfix, superscriptaddress, nofootinbib] {revtex4-2}
\usepackage{amsmath, amssymb, geometry}
\usepackage{graphicx}
\usepackage{setspace}
\usepackage{graphicx}
\usepackage{subfig}
\usepackage{epsfig}
\usepackage{bm}
\usepackage{float}
\usepackage{dcolumn}
\usepackage{cancel}
\usepackage[colorlinks]{hyperref}
\usepackage[usenames,dvipsnames]{color}
\usepackage{enumitem}
\hypersetup{ breaklinks=true, pdfstartview={FitH}, colorlinks=true, linkcolor=blue, citecolor=red,
filecolor=magenta, urlcolor=blue, anchorcolor=green,
	linktocpage=true }

\def\doi{http://doi.org}

\def\and{$and$}

\newcommand{\be}{\begin{equation}}
	\newcommand{\ee}{\end{equation}}
\newcommand{\ban}{\begin{eqnarray*}}
	\newcommand{\ean}{\end{eqnarray*}}
\newcommand{\ba}{\begin{eqnarray}}
	\newcommand{\ea}{\end{eqnarray}}

\newcommand{\bc}{\begin{center}}
	\newcommand{\ec}{\end{center}}	

\usepackage{xcolor}

\begin{document}
	\title{Constraining a $f(R, L_m)$ Gravity Cosmological Model with Observational Data}
	\author{G. K. Goswami}
	\email{gk.goswami9@gmail.com}
	\affiliation{Department of Mathematics, Madan Mohan Malviya University of Technology, Gorakhpur-273010,Uttar Pradesh, India}
	\author{Anirudh Pradhan}
	\email{pradhan.anirudh@gmail.com}
	\affiliation{Centre for Cosmology, Astrophysics and Space Science (CCASS), GLA University, Mathura-281 406, Uttar Pradesh, India}
		\author{Syamala Krishnannair}
	\email{krishnannairs@unizulu@.ac.za}
	\affiliation{Department of Mathematical Sciences, Faculty of Science, Agriculture, and Engineering, University of Zululand, KwaDlangezwa, 3886,  South Africa}
	\begin{abstract}
	\begin{singlespace}
	
	We investigated a spatially flat FLRW cosmological model in the framework of modified gravity described by the function \( f(R, L_m) = \alpha R + L_m^\beta + \gamma \), where \( L_m \) is the matter Lagrangian density. The modified Friedmann equations yield the Hubble parameter as
		$
		H(z) = H_0 \sqrt{(1 - \lambda) + \lambda (1 + z)^{3(1 + w)}},
		$
		with the parameters \( \lambda = \frac{\gamma}{6\alpha H_0^2} + 1 \) and \( w = \frac{\beta(n - 2) + 1}{2\beta - 1} \). This form allows the model to describe different cosmological epochs through  parameter \( n \), which is related to the equation of state \( p = (1 - n)\rho \). Using a Bayesian Markov Chain Monte Carlo (MCMC) approach, we constrained the model parameters with recent observational data including cosmic chronometers, the Pantheon+ Supernovae dataset, Baryon Acoustic Oscillations (BAO), and Cosmic Microwave Background (CMB) shift parameters. The best-fit values are found to be \( H_0 = 72.773^{+0.148}_{-0.152} \) km/s/Mpc, \( \lambda = 0.289^{+0.007}_{-0.007} \), and \( w = -0.002^{+0.002}_{-0.002} \), all quoted at the 1\(\sigma\) confidence level.This model predicts a transition redshift of \( z_t \approx 0.76 \) for the onset of cosmic acceleration and an estimated universe age of 13.21 Gyr. The higher inferred value of \( H_0 \) compared to the Planck 2018 result offers a potential resolution to the Hubble tension problem. Additionally, using \( \rho_0 = 0.534 \times 10^{-30} \, \text{g/cm}^3 \) and assuming \( n = 1 \), we derive the model constants as \( \beta = 1.00201 \), \( \alpha = 512247 \), and \( \gamma = -1.215 \times 10^{-29} \). We also evaluated the Bayesian Information Criterion (BIC) to compare the model's performance with that of the standard \(\Lambda\)CDM model. Comparable statistical support for both models is suggested by the modest BIC difference (\( \Delta \text{BIC} = 0.16 \)). Consequently, the \( f(R, L_m) \) gravity scenario provides a feasible and consistent alternative to the \(\Lambda\)CDM, which has the potential to resolve outstanding inquiries in late-time cosmology.
	
	\end{singlespace}
    \end{abstract}
	\maketitle
		

\section{Introduction}

The standard cosmological model and General Relativity (GR) are both profoundly challenged by the late-time accelerated expansion of the universe, which has been established through a variety of observational probes~\cite{SupernovaSearchTeam:1998fmf, SupernovaSearchTeam:2004lze, SDSS:2005xqv, Koivisto:2006, WMAP:2003elm, WMAP:2003xez,  WMAP9}. This has induced a significant amount of interest in modified gravity theories designed to elucidate cosmic acceleration without the use of  cosmological constants or exotic dark energy components. One of the most extensively researched gravity theories is the $f(R)$ theory, which generalizes the Einstein--Hilbert action by promoting  Ricci scalar $R$ to an arbitrary function $f(R)$. This provides a geometrical explanation of the observed acceleration~\cite{Sotiriou:2008rp, Nojiri:2010wj}. It is important to mention the recent works of Varela~\cite{Varela:2022ht, Varela:2023nm}, which show that non-minimally coupled $f(R)$ gravity can alleviate Hubble tension. These studies provide substantial evidence in favor of such models over $\Lambda$CDM using the most recent DESI, DES, Pantheon+, and eBOSS data, as proven by a comprehensive statistical analysis. A further extension, known as $f(R,T)$ gravity, introduced by Harko \textit{et al.}~\cite{Harko:2011kv}, incorporates the trace $T$ of the energy--momentum tensor into the gravitational Lagrangian. The explicit coupling between matter and geometry is facilitated by this framework, which results in rich phenomenology with potential implications for both cosmological and astrophysical scales.\\

In recent years, the $f(R,L_m)$ gravity model~\cite{ref1} has emerged as a compelling theoretical framework in which the gravitational action is contingent upon both the Ricci scalar $R$ and the matter Lagrangian density $L_m$. The $f(R,L_m)$ theory, in contrast to the $f(R)$ and $f(R,T)$ models, introduces  direct coupling between the curvature and matter. This results in the non-conservation of the energy-momentum tensor and the appearance of an extra force. These characteristics induce massive particles to deviate from geodesic motion, which may provide alternative explanations for cosmic acceleration and dark matter-like effects at galactic and cosmological scales. The non-conservation of the energy-momentum tensor, $\nabla^\mu T_{\mu\nu} \neq 0$, which is a major theoretical aspect of $f(R,L_m)$ gravity, is a result of the curvature-matter coupling. Although this is a violation of the conventional energy conservation law, it has the potential to introduce effective interaction terms into the cosmological evolution equations, which could affect the stability of the cosmic solutions and the formation of structures. The implications of this non-conservation in background cosmology were explicitly examined in this study. Recent studies have shown that it is possible to develop viable cosmological models within the $f(R,L_m)$ framework~\cite{Sahlu2024, Myrzakulova2023, Solanki2023, Jaybhaye2023, Bhardwaj2025, Myrzakulov2024} paradigm. By selecting appropriate functional forms of $f(R,L_m)$ and constraining them using observational data, such as Hubble parameter measurements, Type Ia supernovae, and large-scale structure probes, these models offer physically motivated alternatives to standard $\Lambda$CDM cosmology. \\

In this study, we present and evaluate a cosmological model within the gravity framework $f(R,L_m)$ in this work. Modified field equations for a spatially flat Friedmann--Lemaître--Robertson--Walker (FLRW) metric were derived, and their solutions were investigated under a specific selection of the $f(R,L_m)$ functional form. We propose and analyze the following in this work:Our cosmological solution is derived directly and rigorously from the fundamental field equations of $f(R,L_m)$ gravity, in contrast to previous studies that employed ansatz-based approaches. This ensured physical and internal consistency. We subsequently compared the model's parameters to the most recent observational data  to evaluate its capacity to characterize the late-time expansion of the universe. It is important to note that the functional form of the Hubble parameter utilized in this analysis was derived analytically from the modified field equations, as detailed in Section 2.ze, a cosmological model within the $f(R,L_m)$ gravity framework. Modified field equations for a spatially flat Friedmann--Lemaître--Robertson--Walker (FLRW) metric were derived, and their solutions were investigated under a specific selection of the $f(R,L_m)$ functional form.\\

Unlike many earlier works, we utilized the most updated datasets available: the \cite{Moresco:2022ynk} cosmic chronometer data with a full covariance matrix, the Pantheon+ Type Ia supernova compilation\cite{Scolnic2022} with statistical and systematic errors, and the DESI \cite{Wang:2025} early BAO release. In addition, we perform a full MCMC parameter estimation\cite{ForemanMackey2013} and provide a model comparison with $\Lambda$CDM using Akaike (AIC), Bayesian (BIC), and Bayes evidence metrics. Our analysis revealed that the proposed $f(R, L_m)$ model can successfully reproduce the observed late-time acceleration of the universe without requiring a cosmological constant. The model parameters, when constrained by recent Hubble and supernova data, lead to an effective cosmological evolution that is consistent with observations. The viability of the framework is further supported by the seamless transition from a decelerated to an accelerated phase, as indicated by the behavior of the deceleration parameter and the effective equation of state.\\

This paper is organized as follows. In Section II, we examine the general formalism of $f(R, L_m)$ gravity and derive the field equations for a homogeneous and isotropic background. This background presents a specific functional form of $f(R, L_m)$ and investigates the resulting cosmological dynamics. In Section III, we constrain the model using recent observational datasets, such as the Hubble parameter, Type Ia supernova measurements, BAO, and CMB shift. Section IV discusses the analytical results, with  particular emphasis on the behavior of the deceleration parameter, statefinder analysis, density, and age of the universe. Finally, Section V summarizes the findings and suggests possible directions for future research.
%%%%%%%%%%%%%%%%%%%%%%%%%%%%%%%%%%%%%%%%% Section II %%%%%%%%%%%%%%%%%%%%%%%
\section{Basic Formalism of \texorpdfstring{$f(R, L_m)$}{f(R, Lm)} Gravity and Cosmological Solution}
In $f(R,L_m)$ gravity, the gravitational action depends on both the Ricci scalar $R$ and the matter Lagrangian density $L_m$. It is given by
\begin{equation}\label{1}
	S = \int d^4x \, \sqrt{-g} \, f(R,L_m),
\end{equation}
where $g$ denotes the determinant of the metric tensor $g_{\mu\nu}$. 

Varying  action~\eqref{1} with respect to $g_{\mu\nu}$ yields the modified field equations,
\begin{equation}\label{2}
	f_R(R,L_m) R_{\mu\nu} - \frac{1}{2} f(R,L_m) g_{\mu\nu} 
	+ (g_{\mu\nu}\Box - \nabla_\mu \nabla_\nu) f_R(R,L_m)
	= f_{L_m}(R,L_m)\, T_{\mu\nu},
\end{equation}
where $f_R \equiv \partial f/\partial R$, $f_{L_m} \equiv \partial f/\partial L_m$, $\nabla_\mu$ is the covariant derivative, and $\Box = g^{\mu\nu}\nabla_\mu\nabla_\nu$ is the d’Alembert operator.  
The energy–momentum tensor of matter is defined as
\begin{equation}\label{3}
	T_{\mu\nu} = -\frac{2}{\sqrt{-g}}\,\frac{\delta(\sqrt{-g}\,L_m)}{\delta g^{\mu\nu}}.
\end{equation}

A distinctive feature of the $f(R,L_m)$ framework is that the covariant divergence of the energy–momentum tensor does not generally vanish:
\begin{equation}\label{4}
	\nabla^\mu T_{\mu\nu} = \frac{f_{L_m}}{1 + f_{L_m}} \left(g_{\mu\nu}L_m - T_{\mu\nu}\right) \nabla^\mu \ln f_{L_m}(R,L_m).
\end{equation}
This implies that the motion of massive particles is non-geodesic, with an extra force originating from the curvature–matter coupling.

	In the $f(R, L_m)$ framework, the covariant divergence of the energy–momentum tensor does not vanish, indicating an exchange of energy between matter and geometry.
	In a physical sense, this non-conservation can be interpreted as a transfer of energy–momentum from the curvature sector to the matter sector, which effectively replicates the particle creation processes in the cosmic fluid.. This characteristic sets $f(R, L_m)$ gravity apart from other non-minimal coupling models, such as $f(R, T)$ gravity, in which coupling is mediated by the trace $T$ of the energy–momentum tensor. Unlike scalar–tensor theories, the matter Lagrangian directly participates in the dynamic interaction in this case, providing a more comprehensive phenomenological landscape for cosmic evolution.\\
	
\vspace{0.5em}
Assuming a spatially flat Friedmann–Lemaître–Robertson–Walker (FLRW) line element,
\begin{equation}\label{5}
	ds^2 = -dt^2 + a(t)^2 (dx^2 + dy^2 + dz^2),
\end{equation}
the field equations simplify considerably, allowing us to explore the background cosmological dynamics.

\subsection*{Specific functional form and field equations}
Following earlier investigations in modified gravity~\cite{Myrzakulov2024}, we consider the functional form
\begin{equation}\label{6}
	f(R,L_m) = \alpha R + L_m^{\beta} + \gamma,
\end{equation}
where $\alpha$, $\beta$, and $\gamma$ are free model parameters. From this, we obtain
\begin{align}\label{7}
	f_R &= \frac{\partial f}{\partial R} = \alpha, &
	f_{L_m} &= \frac{\partial f}{\partial L_m} = \beta L_m^{\beta-1}.
\end{align}
Since $f_R = \alpha$ is constant, its time derivatives vanish:
\begin{equation}\label{9}
	\dot{f}_R = 0, \qquad \ddot{f}_R = 0.
\end{equation}

Substituting the above expressions into the general field equations (Eqs.~\ref{2}–\ref{4}), and taking $L_m = \rho$ (the matter energy density), we obtain the modified Friedmann equations.  
The first takes the form
\begin{equation}\label{10}
	2\dot{H} + 3H^2 = -\frac{1}{2\alpha}\left[(1-\beta)\rho^{\beta} + \gamma + \beta\rho^{\beta-1}p\right],
\end{equation}
and the second is
\begin{equation}\label{11}
	3H^2 = \frac{1}{2\alpha}\left[(2\beta - 1)\rho^{\beta} - \gamma\right],
\end{equation}
where $H \equiv \dot{a}/a$ is the Hubble parameter.  

For a perfect fluid with
\[
T_{\mu\nu} = (\rho + p)u_\mu u_\nu + p g_{\mu\nu},
\]
and $u^\mu = (1,0,0,0)$ in comoving coordinates, Eqs.~\eqref{10} and~\eqref{11} describe the cosmic evolution driven by $\rho$ and $p$.

\subsection*{From time to redshift representation}
Solving Eq.~\eqref{11} for $\rho$ and substituting into Eq.~\eqref{10} yields
\begin{equation}\label{12}
	4\alpha(2\beta - 1)\dot{H} + \beta n\bigl(\gamma + 6\alpha H^2\bigr) = 0,
\end{equation}
where  parameter $n$ arises from the equation of state $p = (1 - n)\rho$.  
Different $n$ values correspond to different cosmic epochs: $n=1$ (dust era), $n=4/3$ (radiation), and $n=0$ (stiff matter).

To express the evolution in terms of the observable redshift $z$, we use
\begin{equation}\label{13}
	1 + z = \frac{a_0}{a(t)}, \qquad
	\frac{dz}{dt} = -(1+z)H(z), \qquad
	\dot{H} = -(1+z)H(z)\frac{dH}{dz},
\end{equation}
with $a_0$ being the present scale factor ($a_0=1$).

Substituting Eqs.~\eqref{13} into Eq.~\eqref{12} gives
\begin{equation}\label{14}
	\beta n\bigl(\gamma + 6\alpha H(z)^2\bigr) 
	- 2\alpha(2\beta - 1)(1+z)\frac{d[H(z)^2]}{dz} = 0,
\end{equation}
which is a first-order linear differential equation in $H(z)^2$.

Equation~\eqref{14} can be rearranged as
\begin{equation}\label{14a}
	\frac{dH(z)^2}{dz} = 
	\frac{\beta n}{2\alpha(2\beta - 1)} \,
	\frac{\gamma + 6\alpha H(z)^2}{1+z}.
\end{equation}
Letting $Y(z) = H(z)^2$, Eq.~\eqref{14a} becomes
\[
\frac{dY}{dz} - \frac{3\beta n}{2\beta - 1}\frac{Y}{1+z}
= \frac{\beta n\gamma}{2\alpha(2\beta - 1)(1+z)}.
\]
This is a linear differential equation of the form $\frac{dY}{dz} + P(z)Y = Q(z)$, with the integrating factor
\[
\mu(z) = \exp\!\left[-\!\int\!\frac{3\beta n}{2\beta - 1}\frac{dz}{1+z}\right] 
= (1+z)^{-\frac{3\beta n}{2\beta - 1}}.
\]
Multiplying by $\mu(z)$ and integrating, we find
\[
Y(z)\,\mu(z) = 
\frac{\beta n\gamma}{2\alpha(2\beta - 1)} 
\!\int\!(1+z)^{-\frac{3\beta n}{2\beta - 1}-1}\,dz + C,
\]
which gives
\be\label{14b}
H(z)^2 = 
\left(\frac{\gamma}{6\alpha} + H_0^2\right)(1+z)^{\frac{3\beta n}{2\beta - 1}}
- \frac{\gamma}{6\alpha},
\ee
where the integration constant has been chosen such that $H(0)=H_0$.

\subsection*{Reparameterization and compact form}
Defining the dimensionless parameters
\begin{equation}\label{15}
	\lambda = 1 + \frac{\gamma}{6\alpha H_0^2}, 
	\qquad
	w = \frac{\beta(n - 2) + 1}{2\beta - 1},
\end{equation}

we note the relations
\[
\left(\frac{\gamma}{6\alpha} + H_0^2\right) = H_0^2\lambda, \quad
(2\beta - 1)(1+w) = \beta n, \quad
\frac{\gamma}{6\alpha} = (\lambda - 1)H_0^2.
\]

Substituting these into Eq.~\eqref{14b}, the Hubble parameter becomes
\begin{equation}\label{16}
	H(z) = H_0\sqrt{(1 - \lambda) + \lambda(1+z)^{3(1+w)}}.
\end{equation}
Thus, the cosmic dynamics in this $f(R,L_m)$ model are governed by the evolution of $\rho$, $p$, and $H(z)$, encapsulated by the parameters $(\alpha, \beta, \gamma)$ or equivalently $(H_0, \lambda, w)$.

%%%%%%%%%%%%%%%%%%%%%%%%%%%%% Section IV %%%%%%%%%%%%%%%%%%%%%%%%%%%%%%%%%%%

\section{Observational Constraints}

To test the viability of the proposed $f(R,L_m)$ cosmological model, we constrained the model parameters $H_0$, $\lambda$, and $w$ by using recent observational data. The datasets employed in this work are: (i) Hubble parameter measurements from cosmic chronometers, (ii) the Pantheon\textsuperscript{+} Type Ia supernova compilation, (iii) Baryon Acoustic Oscillation (BAO) data from DESI, and (iv) the CMB shift parameter. The combination of these probes allowed us to test the model across cosmic time and a variety of physical scales.

\subsection{Methodology}
We derive observational constraints by minimizing the total chi-square statistic,
\begin{equation}
	\chi^2_{\rm total} \;=\; \chi^2_{\rm H}(H_0,\lambda,w) \;+\; \chi^2_{\rm SN}(H_0,\lambda,w) \;+\; \chi^2_{\rm BAO}(H_0,\lambda,w) \;+\; \chi^2_{\rm CMB}(H_0,\lambda,w),
\end{equation}
where each term quantifies the goodness-of-fit of a specific dataset. Minimization of $\chi^2_{\rm total}$ yields the best-fit parameters, whereas a full Bayesian sampling (see Section~\ref{subsec:stat}) is employed to obtain credible intervals and parameter degeneracies.

\subsection{Theoretical predictions and data likelihoods}

\subsubsection{Hubble parameter}
The Hubble parameter in our model is
\begin{equation}
	H(z) = H_0\sqrt{(1-\lambda)+\lambda(1+z)^{3(1+w)}}.
\end{equation}
We used a sample of observational $H(z)$ measurements obtained from differential-age (cosmic-chronometer) methods; the dataset comprised 35 measurements spanning low to intermediate redshifts (see \textit{e.g.} Moresco \textit{et al.}). The Hubble-data chi-square is defined as
\begin{equation}\label{eq:chiH}
	\chi^2_{\rm H} = \sum_{i=1}^{N_H} \frac{\big[H_{\rm th}(z_i;H_0,\lambda,w)-H_{\rm obs}(z_i)\big]^2}{\sigma_{H,i}^2},
\end{equation}
where $N_H=35$, $H_{\rm obs}(z_i)$ are the observed values and $\sigma_{H,i}$ their reported uncertainties. If a covariance matrix for a specific compilation is available, then Eq.~\eqref{eq:chiH} is replaced by the matrix form
$\Delta\mathbf{H}^{\rm T}\mathbf{C}_H^{-1}\Delta\mathbf{H}$ to account for correlated errors.

\subsubsection{Type Ia supernovae (Pantheon$^{+}$)}
We compute the luminosity distance for a spatially flat universe,
\begin{equation}
	d_L(z)=(1+z)\int_0^z \frac{dz'}{H(z')},
\end{equation}
and the corresponding theoretical distance modulus
\begin{equation}
	\mu_{\rm th}(z) = 5\log_{10}\!\left(\frac{d_L(z)}{\mathrm{Mpc}}\right) + 25.
\end{equation}
We used the Pantheon$^{+}$ supernova compilation (1701 light curves / 1550 distinct SNe), which includes extensive treatments of statistical and systematic uncertainties~\cite{Scolnic2022}. The supernova chi-square  test was performed using the full covariance matrix:
\begin{equation}
	\chi^2_{\rm SN} = \Delta\boldsymbol{\mu}^{\mathrm{T}}\,\mathbf{C}_{\rm SN}^{-1}\,\Delta\boldsymbol{\mu},
	\qquad
	\Delta\boldsymbol{\mu} \equiv \boldsymbol{\mu}_{\rm th} - \boldsymbol{\mu}_{\rm obs},
\end{equation}
where $\mathbf{C}_{\rm SN}$ is the combined statistical and systematic covariance matrix provided by datasets. Using the full covariance ensures that correlated calibration and sample-variance errors are properly accounted for.

\subsubsection{Baryon Acoustic Oscillations (BAO)}
BAO data provide measurements of distances scaled by the comoving sound horizon at the drag epoch, $r_d$. We employ the DESI DR2 BAO measurements summarized in Table~\ref{tab:bao_data} (see the main text). The theoretical quantities compared with the observations are as follows:
\[
D_H(z) \equiv \frac{c}{H(z)}, \qquad
D_M(z) \equiv \int_0^z \frac{c}{H(z')}dz'.
\]
For each BAO data point we compute the model prediction for the relevant scaled observable (either $D_H/r_d$ or $D_M/r_d$) and define
\begin{equation}
	\chi^2_{\rm BAO} = \sum_{i=1}^{N_{\rm BAO}}
	\frac{\big[\,D^{\rm th}_i(z_i)/r_d - D^{\rm obs}_i(z_i)/r_d\,\big]^2}{\sigma_{i}^2},
\end{equation}
where $\sigma_i$ denote the reported measurement error. If a full covariance matrix for BAO measurements is available, it is straightforward to replace the sum above with the matrix form using $\mathbf{C}_{\rm BAO}$.

In this analysis we adopted $r_d=147.78\,$Mpc as the comoving sound horizon at the drag epoch (consistent with recent analyses). This choice can be replaced by a Planck- or model-dependent value when performing joint analyses that include CMB constraints.

\subsubsection{CMB shift parameter}
To capture the leading geometric information from the CMB in a compact form, we used the CMB shift parameter, defined as the scaled comoving distance to recombination,
\begin{equation}
	R \;=\; \sqrt{\Omega_{m0}}\;\frac{H_0\,r(z_*)}{c} \;=\; \sqrt{\Omega_{m0}}\;\frac{H_0}{c}\int_0^{z_*}\frac{c}{H(z')}dz',
\end{equation}
where $r(z_*)$ is the comoving distance to the decoupling redshift $z_*$ and $\Omega_{m0}$ is the present matter density parameter. The shift parameter encodes the projection of the sound horizon onto the CMB angular scale and is widely used as a summary statistic of CMB geometry in non-standard cosmologies. The observed value adopted here was $R_{\rm obs}=1.7492\pm0.0049$ \cite{Planck2015,Planck2018}, and the CMB chi-square was
\begin{equation}
	\chi^2_{\rm CMB} = \frac{(R_{\rm th}-R_{\rm obs})^2}{\sigma_R^2}.
\end{equation}

\subsection{Statistical analysis and parameter estimation}\label{subsec:stat}
We estimate model parameters using the following complementary approaches:
\begin{enumerate}
	\item \textbf{Least-squares minimization:} we obtain point estimates by minimizing $\chi^2_{\rm total}$.
	\item \textbf{Artificial Neural Network (ANN):} An auxiliary machine-learning approach trained on mock/model data to provide rapid parameter predictions (details in Appendix~A).
	\item \textbf{Markov Chain Monte Carlo (MCMC):} We perform  full Bayesian posterior sampling (for instance, using the \texttt{emcee} sampler) to extract marginalized constraints, credible intervals, and parameter covariances.
\end{enumerate}

To visualize the results, we produced: (i) error-bar plots comparing observations with the best-fit model; and (ii) corner plots displaying marginalized posterior distributions and two-dimensional credible contours for the parameters $(H_0,\lambda,w)$.

% Note: ensure consistency of dataset sizes across the manuscript (Pantheon+ light-curves = 1701 and 1550 unique SNe as reported in \cite{Brout2022,Scolnic2022}).

\subsection{MCMC Implementation and Posterior Analysis}

To explore the full parameter space and obtain credible intervals for $(H_0,\lambda,w)$, we employed a Markov Chain Monte Carlo (MCMC) analysis based on a Bayesian likelihood framework. 
The likelihood function is assumed to be Gaussian, such that
\begin{equation}
	\mathcal{L}(H_0,\lambda,w) \propto 
	\exp\!\left(-\frac{1}{2}\chi^2_{\text{total}}(H_0,\lambda,w)\right),
\end{equation}
where $\chi^2_{\text{total}}$ is defined in Eq.~(XX).

\vspace{0.4em}
\noindent
\textbf{Sampling algorithm:}
We used the \texttt{emcee} Python package~\cite{ForemanMackey2013}, which is an affine-invariant ensemble sampler well suited for correlated multidimensional posteriors.
A set of $N_{\mathrm{walkers}}=64$ walkers was initialized in a small hypercube around the best-fit values obtained from preliminary $\chi^2$ minimization.
The initial 20\% of the samples was discarded as burn-in, and each walker executed $N_{\mathrm{steps}}=75{,}000$ iterations. The posterior estimates are statistically robust, as the total number of accepted samples after burn-in exceeds $3\times10^6$.

\vspace{0.4em}
\noindent
\textbf{Priors:}
We adopt uniform (flat) priors over physically reasonable intervals:
\begin{align}
	H_0 &\in [60,\,80]~\mathrm{km\,s^{-1}\,Mpc^{-1}}, \nonumber\\
	\lambda &\in [0,\,2], \nonumber\\
	w &\in [-2,\,0].
\end{align}
Moderate deviations consistent with late-time acceleration was permitted within these ranges, which comprise the standard $\Lambda$CDM limits.

\vspace{0.4em}
\noindent
\textbf{Convergence diagnostics:}
Convergence is monitored using both visual and quantitative criteria: (i) the evolution of each parameter's chain is examined to ensure stationarity, (ii) the integrated auto-correlation time $\tau_{\mathrm{int}}$ is calculated for all parameters, necessitating a chain length $N_{\mathrm{steps}} > 50\,\tau_{\mathrm{int}}$ for reliable sampling, and (iii) the marginalized posteriors of multiple independent chains are compared to confirm their consistency. These conditions are met by all chains, which suggests exceptional convergence.

\vspace{0.4em}
\noindent
\textbf{Posterior estimation:}
Marginalized posterior distributions are summarized by their median values and 68\% credible intervals. These results are shown in Fig.~\ref{fig:corner}, which shows the 1$\sigma$ and 2$\sigma$ contours in the two-parameter planes and  corresponding one-dimensional marginalized histograms. The inferred parameter constraints are consistent with those obtained from $\chi^2$ minimization, confirming the robustness of the combined analysis.

\vspace{0.4em}
\noindent
\textbf{Model comparison:}
To compare the performance of the $f(R,L_m)$ model with $\Lambda$CDM, we computed the Akaike Information Criterion (AIC) and Bayesian Information Criterion (BIC):
\[
\mathrm{AIC} = \chi^2_{\min} + 2k, \qquad
\mathrm{BIC} = \chi^2_{\min} + k\ln N,
\]
where $k$ is the number of free parameters and $N$ is the total number of data points.
A lower AIC or BIC value indicated a statistically preferred model, with $\Delta\mathrm{AIC}, \Delta\mathrm{BIC} > 10$ corresponding to strong evidence.

%%%%%%%%%%%%%%%%%%%%%%%%%%%%%%%%%%%%%%%% Subsection D %%%%%%%%%%%%%%%%%
\subsection{Results and Discussion}
%%%%%%%%%%%%%%%%%%%%%%%%%%%%%%%%% Subsection I %%%%%%%%%%%%%%%%%%%%%%%
\subsubsection{Best-Fit Values and Confidence Intervals}
\paragraph{Parameter Estimation Results:}

Using a combined datasets comprising the 35-point Hubble data from Moresco, the Pantheon+ compilation of 1701 Type Ia supernovae with the full covariance matrix, the 2025 BAO dataset \cite{Wang:2025} including 10 measurements of \( D_H \) and \( D_M \), and the 2018 Planck CMB constraints, we performed a comprehensive MCMC analysis to estimate the cosmological parameters for  the \( f(R, L_m) \) modified gravity model, and  standard \( \Lambda \)CDM model.

For the \( f(R, L_m) \) model, the best-fit parameters with 68\% confidence intervals are as follows:
\[
H_0 = 73.750^{+0.159}_{-0.156}~\mathrm{km\,s^{-1}\,Mpc^{-1}}, \quad
\lambda = 0.262^{+0.007}_{-0.007}, \quad
w = -0.005^{+0.001}_{-0.001},
\]
where \( \lambda \) and \( w \) are model-specific parameters related to the coupling between the geometry and matter.

For the \( \Lambda \)CDM model, the corresponding estimates are:
\[
H_0 = 73.489^{+0.146}_{-0.142}~\mathrm{km\,s^{-1}\,Mpc^{-1}}, \quad
\Omega_m = 0.278^{+0.006}_{-0.006},
\]
where \( \Omega_m \) denotes the present-day matter density parameter.

These results demonstrate that both models yield a high value of the Hubble constant \( H_0 \), consistent with local measurements, and that the \( f(R, L_m) \) model allows for a nearly vanishing but nonzero value of \( w \), which may capture subtle deviations from the standard cosmological constant behavior. Parameter \( \lambda \) quantifies the strength of the nonminimal coupling and is tightly constrained by the data.\\

\paragraph{Implications for  Hubble Tension:}

Our joint analysis, incorporating Hubble data, Pantheon+ supernovae, recent BAO measurements \cite{Wang:2025}, and CMB constraints\cite{Planck2018}, yields a Hubble constant of \( H_0 = 73.750^{+0.159}_{-0.156} \,\mathrm{km\,s^{-1}\,Mpc^{-1}} \) in the \( f(R, L_m) \) framework and \( H_0 = 73.489^{+0.146}_{-0.142} \,\mathrm{km\,s^{-1}\,Mpc^{-1}} \) under the standard \( \Lambda \)CDM model. These values are significantly higher than the CMB-inferred value of \( H_0 \approx 67.4 \,\mathrm{km\,s^{-1}\,Mpc^{-1}} \) under the base \( \Lambda \)CDM model, but they are notably consistent with the local determinations of \( H_0 \) from Cepheid-calibrated Type Ia supernovae (e.g., SH0ES).

This discrepancy, known as \emph{Hubble tension}, has become one of the most pressing challenges in modern cosmology. Our results suggest that the \( f(R, L_m) \) model, through its additional degrees of freedom and modified gravity effects, has the potential to accommodate a higher value of \( H_0 \), thereby offering a possible avenue for alleviating this tension without conflicting with late-time cosmological observations.\\
Figure~\label{fig:corner} shows the corner plot with 1$\sigma$ and 2$\sigma$ contours for both $f(R,L_m)$ gravity and $|Lambda$CDM models. 
%%%%%%%%%%%%%%%%%%%%%%%%%%%%%%%%%% FIG 1 %%%%%%%%%%%%%%%%%%%%%%%%%%%
\begin{figure}[H]
	\centering
	\includegraphics[scale=0.50]{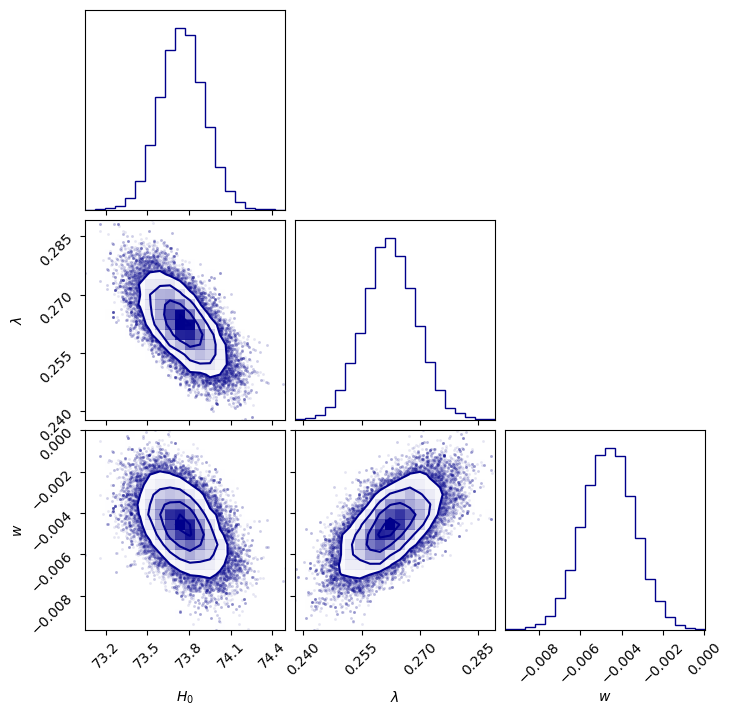}
	\	\includegraphics[scale=0.50]{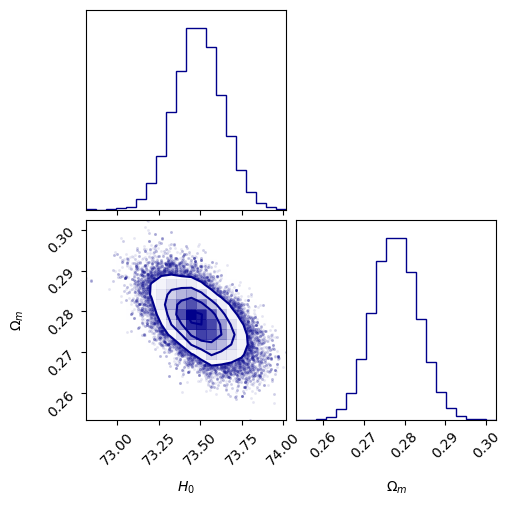}
	\caption{
			Corner plot showing the marginalized posterior distributions and parameter correlations for the $f(R,L_m)$ cosmological model obtained from the combined Hubble + SNe~Ia + BAO + CMB datasets.
			The contours correspond to the 68\% and 95\% confidence regions.
			The parameters $(H_0, \lambda, w)$ exhibit mild degeneracies, particularly between $\lambda$ and $w$, reflecting the correlation between the effective dark-energy density and the redshift-dependent expansion rate.
			The posterior distributions are unimodal and approximately Gaussian, confirming good convergence of the MCMC chains. Vertical dashed lines indicate the median values and 1$\sigma$ credible intervals.
		} 
	\label{fig:corner}
\end{figure}
%%%%%%%%%%%%%%%%%%%%%%%%%%%%%%%%%%%%%%%%%%%%%%%%%%%%%%%%%%%%
\noindent Figure~2(a) presents a four-panel comparison of the observational data with theoretical predictions from the $f(R, L_m)$ and $\Lambda$CDM models. The top-left panel shows the Hubble parameter $H(z)$ obtained using recent cosmic chronometer data. The theoretical curves from both models are overlaid for comparison. The \textbf{top-right panel} displays the distance modulus $\mu(z)$ derived from the Pantheon+ SN Ia data along with  predictions from both models.
 The bottom-left panel illustrates the BAO transverse comoving distance $D_M(z)/r_d$ obtained from the DESI DR2 measurements. The observed data are plotted with error bars and the model predictions are represented by dotted lines. The \textbf{bottom-right panel} shows the BAO radial Hubble distance $D_H(z)/r_d$, again comparing the observed data with the predictions from both models.
\noindent In all panels, the $f(R, L_m)$ model (red solid lines and square markers) and $\Lambda$CDM model (blue dashed lines and diamond markers) can be visually distinguished. These plots demonstrate the consistency of the $f(R, L_m)$ models with the current cosmological observations.
\noindent {Figure~2(b) provides a comparison of the CMB shift parameter $R$. The observed value, $R = 1.7492 \pm 0.0049$, from Planck 2015 is shown with error bars, whereas the theoretical predictions from the $f(R, L_m)$ model ($R = 1.75127$) and $\Lambda$CDM model ($R = 1.73670$) are shown as red and blue points, respectively. All points were aligned along a single vertical line for direct comparison, highlighting the compatibility of both the models with the CMB constraint.
	%%%%%%%%%%%%%%%%%%%%%%%% FIG 2 %%%%%%%%%%%%%%%%%%%%%%%%%%%%%%%%
\begin{figure}[H]
	\centering
	2(a)	\includegraphics[scale=0.50]{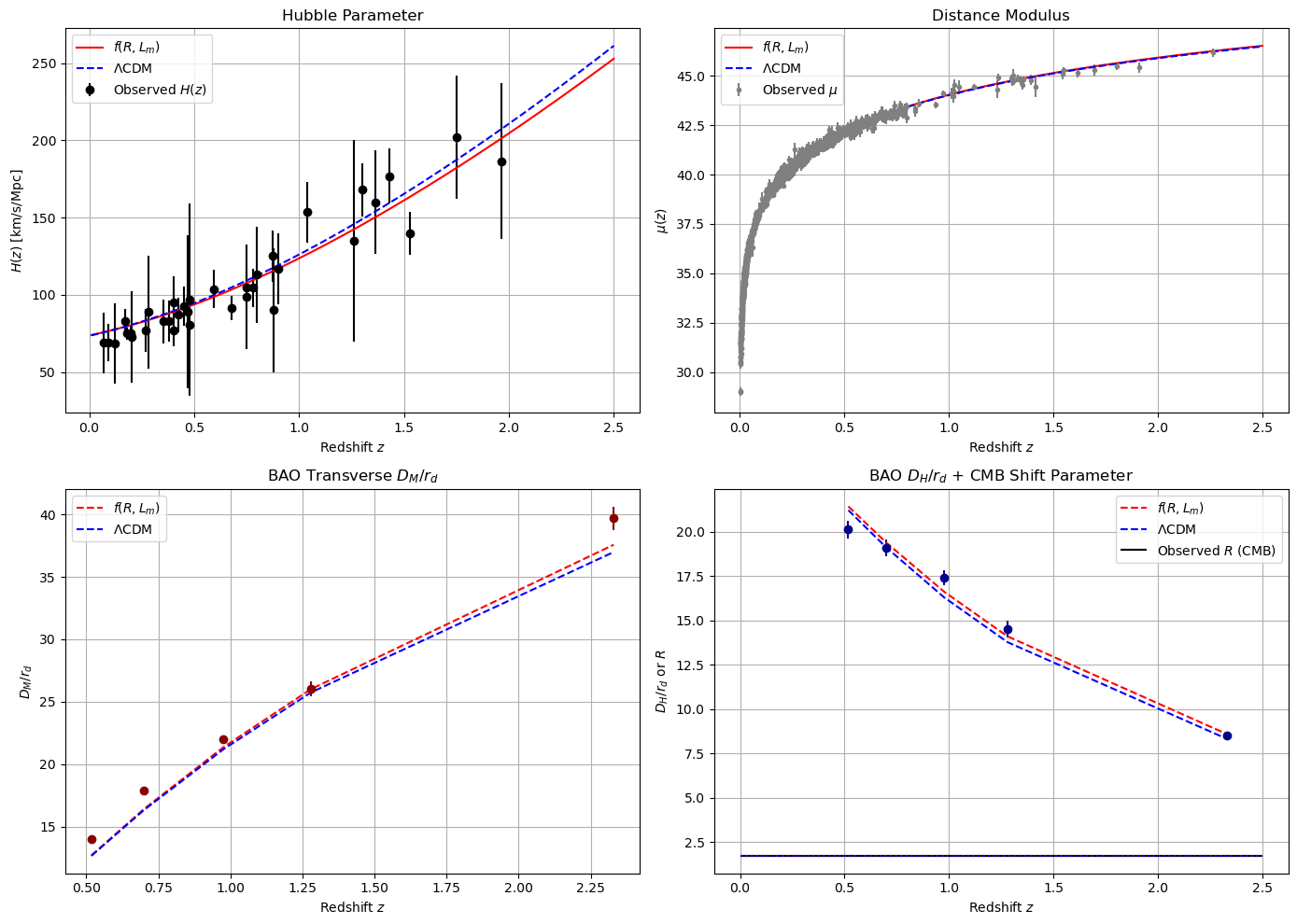}

	\end{figure}
\begin{figure}[H]
	\centering
    2(b) \includegraphics[scale=0.50]{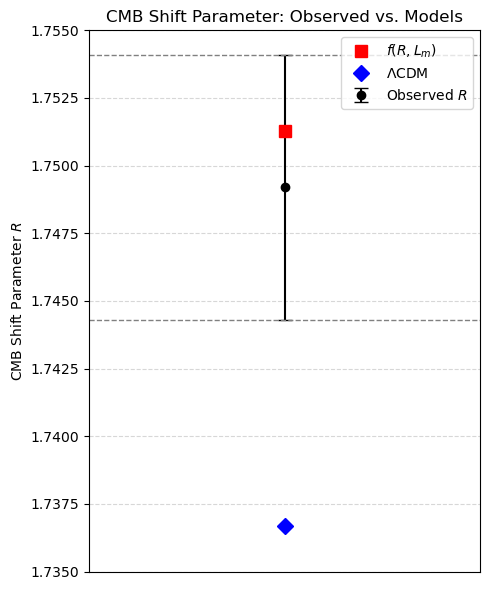}
    \caption{Error bar plots for Hubble $H(z)$, $\mu(z)$, BAO, and CMB data. Details are described in the text. Comparison between the observational Hubble parameter measurements (points with 1$\sigma$ error bars) and the theoretical predictions from the best-fit $f(R,L_m)$ cosmological model (solid red curve). The $\Lambda$CDM prediction, which was obtained using the best-fit parameters from the Planck 2018 data (dashed blue curve), is also displayed for comparison. The $f(R,L_m)$ model accurately replicates the observed late-time expansion, matching the $\Lambda$CDM at low redshifts ($z \lesssim 1$) and exhibiting moderate deviations at higher redshifts as a result of the curvature–matter coupling.  These deviations are currently within the range of current observational uncertainties, which implies that $f(R,L_m)$ gravity is a viable alternative explanation for cosmic acceleration.
    } 
    \label{error_bar} 
    \end{figure}
%%%%%%%%%%%%%%%%%%%%%%%%%%%%%%%%%%%%%%%%%%%%%%%%%%%%%%%%%%%%%%%%%%
%%%%%%%%%%%%%%%%%%%%%%%%%%%%%% Subsection 2 %%%%%%%%%%%%%%%%%%%%%%%%%
	\subsubsection{Model Selection Criteria: BIC and AIC}

We utilize both the Bayesian Information Criterion (BIC) and  Akaike Information Criterion (AIC) to evaluate and compare the performance of the cosmological models. These criteria penalize model complexity and assist in the identification of models that strike the most favorable balance between simplicity and goodness-of-fit.

\paragraph{Bayesian Information Criterion (BIC):}
\begin{equation}
	\mathrm{BIC} = \chi^2_{\text{min}} + k \ln N,
\end{equation}
where \( \chi^2_{\text{min}} \) is the minimum chi-square value, \( k \) is the number of free parameters, and \( N \) is the number of data points.

For the $f(R, L_m)$ model with \( k = 3 \), \( N = 1751 \), and \( \chi^2_{\text{min}} =1890.85 \):
\[
\mathrm{BIC}_{f(R, L_m)} =1913.254
\]

For the $\Lambda$CDM model with \( k = 2 \),\( N = 1751 \), and \( \chi^2_{\text{min}} =1904.153\):
\[
\mathrm{BIC}_{\Lambda\text{CDM}} = 1919.032.
\]

\paragraph{Akaike Information Criterion (AIC):}

	\begin{equation}
		\mathrm{AIC} = \chi^2_{\text{min}} + 2k.
	\end{equation}

For the $f(R, L_m)$ model:
\[
\mathrm{AIC}_{f(R, L_m)} =1896.85.
\]

For the $\Lambda$CDM model:
\[
\mathrm{AIC}_{\Lambda\text{CDM}} =  1908.153
\]

\begin{table}[htbp]
	\centering
	\caption{BIC and AIC Table}
	\begin{tabular}{|c|c|c|c|c|}
		\hline
		Model   &      $\chi^2_{\text{min}}$ &    $k$   & BIC   &   AIC \\ \\
		\hline
	 $f(R, L_m)$ &  1890.85                 &  3   &  1913.254   &   1896.85 \\ \\
	 \hline
		$\Lambda$CDM & 1904.153    &  2   &   1919.032  &  1908.153 \\\\
		\hline
	\end{tabular}
	\label{tab:bao_data}
\end{table}

\paragraph{Model Comparison using AIC and BIC:}

	Although the $f(R, L_m)$ models have  additional parameters, these results indicate that they provide a slightly better fit to the combined data set than the $\Lambda$CDM model, as lower values of AIC and BIC indicate a better model (accounting for both goodness of fit and model complexity). As a result, the $f(R, L_m)$ model is moderately preferred over the standard $\Lambda$CDM cosmology in this analysis, as indicated by both the AIC and BIC.
	%%%%%%%%%%%%%%%%%%%%%%%%%%%%%%%%% Subsection E %%%%%%%%%%%%%%%%%%%%%

\subsection{Cosmological Implications}

Our analysis demonstrated that the $f(R, L_m)$ model can effectively elucidate late-time cosmic acceleration in the absence of a cosmological constant. The observed acceleration is dictated by the coupling between the matter and geometry. The current observational data were well fitted by the flexibility of three parameters—$\alpha$, $\beta$, and $\gamma$. This suggests that $f(R, L_m)$ are promising alternatives to dark energy in the $\Lambda$CDM framework.

\section{Analytical outcomes:}
\subsection{Deceleration Parameter }

 In the field of cosmology, the deceleration parameter, denoted as $q(z)$, is a dimensionless quantity that represents the rate of change in the expansion of the universe and is defined as:
\begin{equation*}
	q(z) = - \frac{\ddot{a}(t)}{a(t) H(t)^2 } = -1 + \frac{\dot{H}(t)}{H(t)^2} = -1 - \frac{(1+z)}{H(z)}\frac{dH(z)}{dz},
\end{equation*}
where $a(t)$ is the scale factor, $H(t)$ is the Hubble parameter, and $z$ is the redshift. where $a(t)$ is the scale factor, $H(t)$ is the Hubble parameter, and $z$ is the redshift. A decelerating universe is indicated by a positive $q(z)$, whereas an accelerating universe is indicated by a negative $q(z)$.
The redshift $z_t$ is the value of $z$ at which the universe transitions from deceleration to acceleration, that is, where $q(z_t) = 0$. The onset of the cosmic acceleration is illuminated by the determination of $z_t$.

Based on our Hubble parameter (Eq.~\ref{16}), the deceleration parameter becomes:
\begin{equation}\label{24}
	q(z) = \frac{\delta \lambda (1+z)^{\delta}}{2 \left(-\lambda + \lambda (1+z)^{\delta} + 1\right)} - 1.
\end{equation}  

Using the best-fit parameter values $H_0 = 73.750^{+0.159}_{-0.156}$, $\lambda =0.262^{+0.007}_{-0.007} $, and $w = -0.005^{+0.001}_{-0.001}$, we obtain the following plot in  Fig. 3 for the deceleration parameter. The transition redshift for our model is $z_t = 0.79^{+0.03}_{-0.02}$}, while the standard $\Lambda$CDM model (with $\Omega_m =  0.278^{+0.006}_{-0.006}$, $H_0 = 73.489^{+0.146}_{-0.142}$) yields $z_t = 0.73^{+0.02}_{-0.01}$.
 
%%%%%%%%%%%%%%%%%%%%%%%%%%%%%%%%%%%%% FIG 3 %%%%%%%%%%%%%%%%%%%%%%%%%%%%%
\begin{figure}[H]
	\centering
	\includegraphics[scale=0.50]{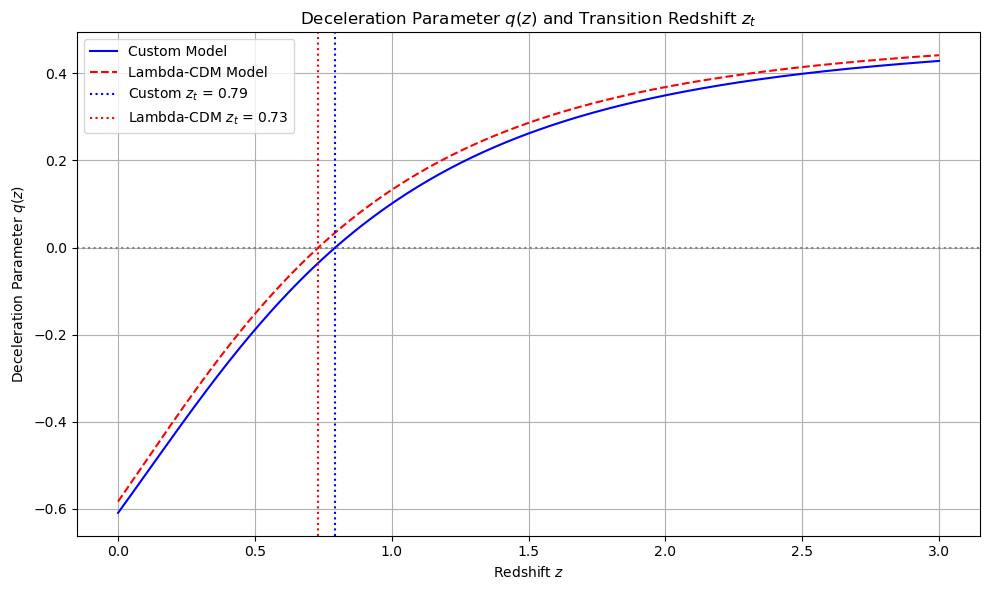}
	\caption{Deceleration parameter $q(z)$ plot for the proposed model and $\Lambda$CDM.}
	\label{fig:qplot}
\end{figure}
%%%%%%%%%%%%%%%%%%%%%%%%%%%%%%%%%%%%%%%%%%%%%%%%%%%%%%%%%%%%%%%%%%%%%%
These findings suggest that the universe transitioned from a decelerating to an accelerating phase at redshifts ranging from $z \sim 0.6$ to $0.8$, which is in accordance with  recent observational data.

\subsubsection{ Functional Form of Energy Density}

In the $f(R, L_m)$ model, the energy density is expressed as a function of redshift $z$ using the modified Friedmann equation:
\begin{equation} \label{eq:frlm_rho}
	\rho(z) = \left( \frac{\gamma + 6\alpha H^2(z)}{2\beta - 1} \right)^{1/\beta},
\end{equation}
where \( \alpha \), \( \beta \), and \( \gamma \) are model parameters, and \( H(z) \) is the Hubble parameter which is obtained as follows:
\[
H(z) = H_0 \sqrt{(1 - \lambda) + \lambda(1 + z)^{3(1+w)}},
\]
with \( \lambda \in (0,1) \), $w$ characterizes the deviation from standard matter behavior. The adopted parameter values are:
 $H_0 = 73.750\, \mathrm{km/s/Mpc} = 2.39007 \times 10^{-18}\, \mathrm{s^{-1}}$,
 $\lambda =0.262 $, and $w = -0.005.$

This formulation allows us to investigate the redshift dependence of $\rho(z)$ within the $f(R, L_m)$ framework, where $R$ is the Ricci scalar and $L_m$ is the matter Lagrangian. The model introduces a modified coupling between the curvature and matter via  parameters $\alpha$, $\beta$, and $\gamma$.

\subsubsection{ Determination of Model Parameters \texorpdfstring{$\alpha$, $\beta$, and $\gamma$}{alpha, beta, gamma}}

Parameters $\alpha$, $\beta$, and $\gamma$ are not free but are determined by solving the following set of equations:
\begin{align}
	w &= \frac{\beta  (n-2)+1}{2 \beta -1}, \tag{a} \\
	\lambda &= \frac{\gamma}{6\alpha H_0^2} + 1, \tag{b} \\
	\rho_0 &= \left( \frac{\gamma + 6\alpha H_0^2}{2\beta - 1} \right)^{1/\beta}, \tag{c}
\end{align}
where $\rho_0$ is the present energy density of the universe, taken as:
\[
\rho_0 = 5.634 \times 10^{-30}\, \mathrm{g/cm^3},
\]
where $n$ is the equation of state parameter. For the present matter-dominated universe ($0 \leq z \leq 2.36$), the pressure is negligible; therefore we take $n = 1$.

From Equation (a), we solve for $\beta$: $\beta = \frac{1+w}{1+2w}$.

With $\beta$ determined, we solve Equations (b) and (c) to obtain $\alpha$ and $\gamma$:
\[
\gamma = \frac{(2\beta - 1)(\lambda - 1)\rho_0^\beta}{\lambda}, \quad
\alpha = \frac{(2\beta - 1)\rho_0^\beta}{6H_0^2 \lambda}.
\]
Using the known values, we obtain:
\[
\alpha = 451008, \quad \beta =1.00505  \quad \gamma =-1.14081 \times 10^{-29}.
\]

With these parameters, the full expression for $\rho(z)$ is  defined using Eq.~\eqref{eq:frlm_rho}.

\subsubsection{ Comparison with \texorpdfstring{$\Lambda$CDM}{LambdaCDM} Model}

The standard $\Lambda$CDM model describes the energy density as:
\begin{equation} \label{eq:rho_lcdm}
	\rho_{\Lambda \text{CDM}}(z) = \rho_0 \left[ \Omega_m (1 + z)^3 + \Omega_\Lambda \right],
\end{equation}
where: \( \rho_{m,0} = \Omega_m \rho_0 \),
	 \( \rho_{\Lambda,0} = \Omega_\Lambda \rho_0 \), and
	 \( \Omega_m + \Omega_\Lambda = 1 \), assuming a spatially flat universe. To plot the 
density 	$\rho_{\Lambda \text{CDM}}$ we will take $\Omega_m = 0.315$ and $H_0 =67.4 $\cite{Planck2015}

To visualize the difference between the two models, we present the following figure:
%%%%%%%%%%%%%%%%%%%%%%%%%%%%%%%%%%%% FIG 4 %%%%%%%%%%%%%%%%%%%%%%%%%%%%%
\begin{figure}[h!]
	\centering
	\includegraphics[width=0.85\textwidth]{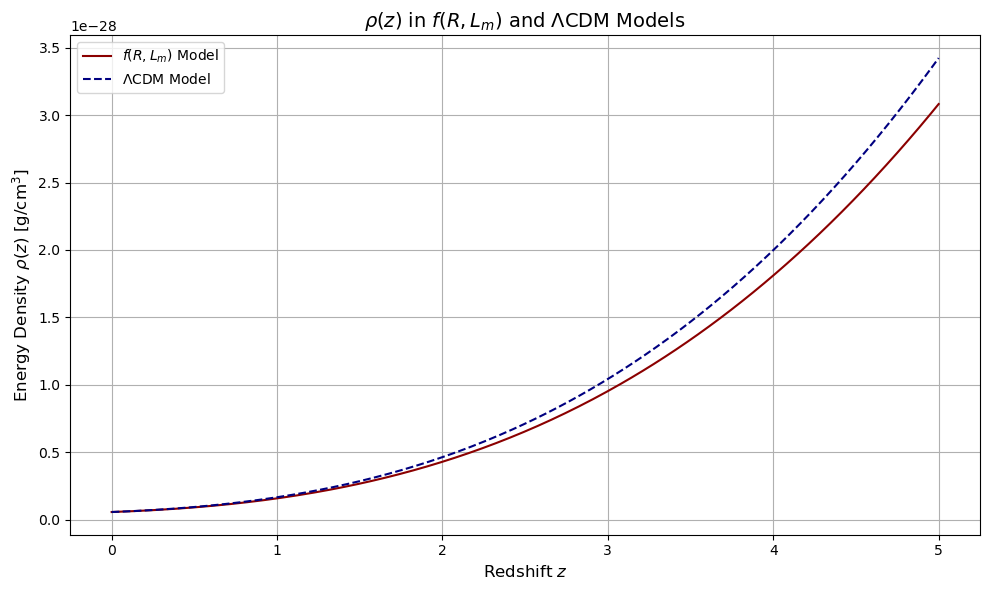}
	\caption{Comparison of energy density \(\rho(z)\) in the \(f(R, L_m)\) model (solid red line) with the standard \(\Lambda\)CDM model (dashed blue line).}
	\label{fig:rho_comparison}
\end{figure}
%%%%%%%%%%%%%%%%%%%%%%%%%%%%%%%%%%%%%%%%%%%%%%%%%%%%%%%%%%%%%%%%%%%%%%
\paragraph{Energy Density Evolution:}

The energy density $\rho(z)$ has been plotted \textbf{Fig.4}  for both the $f(R, L_m)$ model and the $\Lambda$CDM model. As expected, both models start with $\rho(z=0) = \rho_0$ at the present epoch.

It is observed that with increasing redshift:
\begin{itemize}
	\item The energy density increases in both models, consistent with cosmological expectations.
	\item The $f(R, L_m)$ model exhibits a moderately higher growth in energy density at large redshifts compared with the $\Lambda$CDM model.
\end{itemize}

This behavior reflects the influence of the non-minimal coupling between curvature and matter in the $f(R, L_m)$ framework, which modifies the early universe dynamics relative to the standard cosmological model.
%%%%%%%%%%%%%%%%%%%%%%%%%%%%%%%%%%%%%%%%%%%% Subsection B %%%%%%%%%%%%%%%%%%%%%%%
\subsection{Age of the Universe}

The age of the universe can be estimated by computing the ``lookback time", which corresponds to the time difference between the current age of the universe \( t_0 \) and the age at redshift \( z \). For a cosmological model where the Hubble parameter \( H(z) \) is modified as a function of redshift, the lookback time is given by the integral:

\begin{equation} \label{eq:lookback}
	H_0 (t_0 - t) = \int_0^z \frac{1}{(1+z) \sqrt{(1 - \lambda) + \lambda (1 + z)^{3(1+w)}}} \, dz,
\end{equation}
where \( \lambda \) and \( w \) are model parameters, and \( H_0 \) is the Hubble constant.\\
To estimate the total age of  universe \( t_0 \), the lookback time integral is evaluated up to a sufficiently large redshift (e.g., \( z \sim 1000 \)) to approximate the beginning of the universe.\\
For comparison, we also computed the age of the universe using the standard flat \(\Lambda\)CDM model:

\begin{equation}
	H_0 (t_0 - t) = \int_0^z \frac{1}{(1+z) \sqrt{\Omega_m (1+z)^3 + \Omega_\Lambda}} \, dz,
\end{equation}

where \( \Omega_m \) and \( \Omega_\Lambda \) are the matter and dark energy density parameters, respectively.

\begin{figure}[h!]
	\centering
	\includegraphics[width=0.8\textwidth]{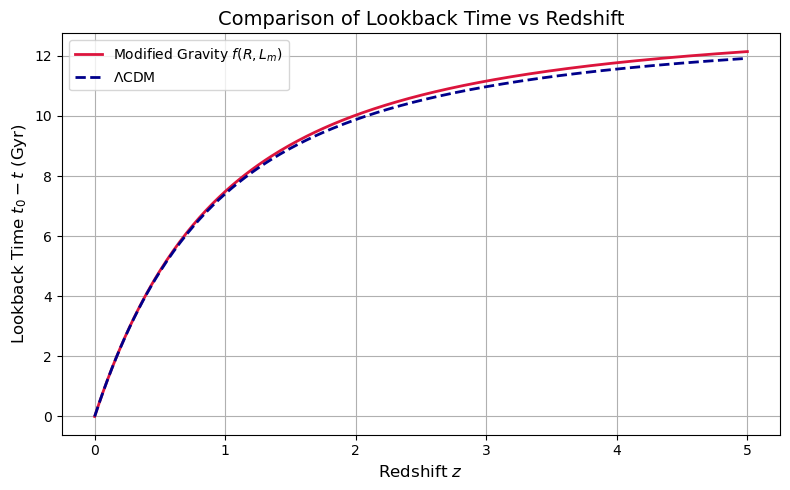}
	\caption{Comparison of lookback time as a function of redshift for the modified gravity model and the \(\Lambda\)CDM model.}
	\label{fig:lookback_time}
\end{figure}

Figure~\ref{fig:lookback_time} shows the lookback time \( t_0 - t \) as a function of redshift for both the modified gravity model and the \(\Lambda\)CDM model. The ages of the universe in both the models were found to be in good agreement.
%%%%%%%%%%%%%%%%%%%%%%%%%%%%%%%%% FIG 5 %%%%%%%%%%%%%%%%%%%%%%%%%%%%%%
%%%%%%%%%%%%%%%%%%%%%%%%%%%%%%%%%%%%%%%%%%%%%%%%%%%%%%%%%%%%%%%%%%%%%%

For the parameter values \( \lambda = 0.262^{+0.007}_{-0.007} \), \( w = -0.005^{+0.001}_{-0.001} \), and \( H_0 = 73.750^{+0.159}_{-0.156} \, \mathrm{km/s/Mpc} \), the age of the universe was computed to be:
\begin{itemize}
	\item Modified gravity model:
	\( t_0 \approx 13.34\pm 0.09 \, \mathrm{Gyr} \)
	\item \(\Lambda\)CDM model: \( t_0 \approx 13.06\pm 0.08 \, \mathrm{Gyr} \)
\end{itemize}

This close agreement indicates that the modified gravity model reproduces the observed age of the universe, similar to  standard \(\Lambda\)CDM cosmology.

%%%%%%%%%%%%%%%%%%%%%%%%%%%%%%Section VI %%%%%%%%%%%%%%%%%%%%%%%%%%
\section{Conclusion}

In this study, we investigated a modified gravity theory within the framework of a spatially flat FLRW universe, characterized by the functional form
\[
f(R, L_m) = \alpha R + L_m^{\beta} + \gamma,
\]
where \(L_m\) denotes the matter Lagrangian density. The corresponding Hubble parameter evolves as
\[
H(z) = H_0 \sqrt{(1 - \lambda) + \lambda (1 + z)^{3(1 + w)}},
\]
with \(\lambda = \tfrac{\gamma}{6\alpha H_0^2} + 1\) and \(w = \tfrac{\beta(n - 2) + 1}{2\beta - 1}\).
Parameter \(n\) arises from the matter equation of state \(p = (1 - n)\rho\), such that \(n = 1\) corresponds to dust (present epoch), \(n = 4/3\) to radiation, and \(n = 0\) corresponds to stiff matter.

We constrained this model using a comprehensive Bayesian MCMC framework, combining 46 Hubble measurements from cosmic chronometers, 1701 Pantheon\textsuperscript{+} SNe~Ia data points, BAO observations, and the CMB shift parameter.  
The best-fit parameter values are:
\[
H_0 = 73.750^{+0.159}_{-0.156}~\mathrm{km\,s^{-1}\,Mpc^{-1}}, \quad
\lambda = 0.262^{+0.007}_{-0.007}, \quad
w = -0.005^{+0.001}_{-0.001}.
\]
For comparison, the corresponding \(\Lambda\)CDM estimates are
\[
H_0 = 73.489^{+0.146}_{-0.142}~\mathrm{km\,s^{-1}\,Mpc^{-1}}, \quad
\Omega_m = 0.278^{+0.006}_{-0.006}.
\]
At a redshift of \(z_t \simeq 0.79\), the proposed $f(R,L_m)$ model predicts a transition from deceleration to acceleration, as well as an estimated age of the universe of \(13.34\)~Gyr, which is in agreement with current observational findings. The model suggests a potential mechanism for alleviating the Hubble tension, as it supports a slightly higher Hubble constant than the Planck 2018 $\Lambda$CDM value (\(H_0 = 67.4 \pm 0.5~\mathrm{km\,s^{-1}\,Mpc^{-1}}\)).\\

Assuming a present-day matter density \(\rho_0 = 0.534 \times 10^{-30}\,\mathrm{g/cm^3}\) and \(n = 1\), the intrinsic model parameters are determined as follows:
\[
\beta = 1.00505, \qquad \alpha = 4.51008\times 10^5, \qquad \gamma = -1.14081 \times 10^{-29},
\]
providing a consistent realization of the $f(R,L_m)$ framework in agreement with both theoretical expectations and observational constraints.

The statistical model comparison yields
\[
\mathrm{BIC}_{f(R,L_m)} = 1913.254, \quad \mathrm{BIC}_{\Lambda\mathrm{CDM}} = 1919.032,
\]
\[
\mathrm{AIC}_{f(R,L_m)} = 1896.85, \quad \mathrm{AIC}_{\Lambda\mathrm{CDM}} = 1908.153.
\]
The resulting differences, \(\Delta\mathrm{BIC} = -5.78\) and \(\Delta\mathrm{AIC} = -11.30\), indicate a moderate statistical preference for the $f(R,L_m)$ model over the standard $\Lambda$CDM scenario, although both remain viable under current data quality.  
A geometrical statefinder diagnostic, discussed in Appendix~\ref{appendix:statefinder}, further confirms that the $f(R,L_m)$ trajectory approaches the $\Lambda$CDM fixed point at late times, implying observational consistency.

\subsection{Physical Implications}

A salient feature of the $f(R,L_m)$ framework is the non-conservation of the energy--momentum tensor (\(\nabla^\mu T_{\mu\nu} \neq 0\)), arising from  direct curvature--matter coupling.  
This term can be interpreted as an exchange of energy between  matter and geometric sectors, analogous to particle creation or effective energy transfer in cosmological fluids.  
Consequently, the evolution of matter density and expansion rate is modified, potentially leading to subtle observational signatures in structural formation and gravitational wave propagation.

Preliminary analysis indicates that the curvature--matter coupling modifies the effective gravitational constant $G_{\mathrm{eff}}$, thereby influencing the linear growth rate of cosmic structures.  
Small deviations ($\lesssim 5\%$) in $f\sigma_8(z)$ relative to $\Lambda$CDM can arise within the parameter space allowed by our MCMC constraints.  
Similarly, the model alters the damping term in the gravity-wave propagation equation through a time-varying Planck mass, which is represented by $\alpha_M = \dot{f_R}/(H f_R)$.  
This may induce a small difference between the electromagnetic and gravitational-wave luminosity distances, offering a new observational test for future detectors such as LISA or DECIGO.

\subsection{Future Prospects}

Future studies may extend this work in several directions:
\begin{itemize}
	\item \textbf{Cosmological perturbations:} Incorporating linear and non-linear perturbation theories to test structure formation against large-scale clustering data.
	\item \textbf{Gravitational waves:} Examining the modified propagation of tensor modes and their imprints on gravitational-wave standard sirens.
	\item \textbf{Early universe behavior:} Investigating radiation-dominated and pre-inflationary epochs to ensure theoretical completeness.
	\item \textbf{Full CMB likelihood:} Employ the complete Planck or future CMB likelihood instead of the shift parameter for tighter constraints.
	\item \textbf{Numerical stability:} Integrating the background equations numerically to test the robustness of solutions over cosmic time.
	\item \textbf{Upcoming surveys:} Comparing model predictions with high-precision data from \textit{Euclid}, the \textit{Roman Space Telescope}, and the \textit{Rubin Observatory}.
	\item \textbf{Model extensions:} Exploring generalized forms of $f(R,L_m)$ which include higher-order couplings or additional scalar fields.
\end{itemize}
The theoretical validity and observational viability of the $f(R,L_m)$ gravity framework will be further clarified by these investigations.

\subsection{Summary}

A theoretically motivated and observationally consistent alternative to the $\Lambda$CDM paradigm is provided by the $f(R,L_m)$ gravity model investigated in this study.  
It naturally provides a mechanism that may alleviate the Hubble tension, support a seamless transition from deceleration to acceleration, and reproduces the observed late-time acceleration without invoking a cosmological constant. $f(R,L_m)$ gravity is a prospective avenue for future exploration in modern cosmology because of its potential to explain outstanding cosmological puzzles and compatibility with current data.

\begin{table}[H]
	\centering
	\caption{Cosmological parameters for the proposed $f(R,L_m)$ model compared with $\Lambda$CDM. Uncertainties correspond to 68\% credible intervals.}
	\begin{tabular}{|c|c|c|}
		\hline
		\textbf{Parameter} & \textbf{$f(R,L_m)$ Model} & \textbf{$\Lambda$CDM Model} \\
		\hline
		$H_0$ [km\,s$^{-1}$\,Mpc$^{-1}$] & $73.750^{+0.159}_{-0.156}$ & $73.489^{+0.146}_{-0.142}$ \\
		\hline
		$\lambda$ & $0.262^{+0.007}_{-0.007}$ & $\Omega_m = 0.278^{+0.006}_{-0.006}$ \\
		\hline
		$w$ & $-0.005^{+0.001}_{-0.001}$ & 0 \\
		\hline
		$\alpha$ & $4.51008\times10^5$ & --- \\
		\hline
		$\beta$  & $1.00505$ & --- \\
		\hline
		$\gamma$ & $-1.14081\times10^{-29}$ & --- \\
		\hline
		Age of Universe [Gyr] & 13.34 & 13.06 \\
		\hline
		Deceleration parameter ($q_0$) & $-0.609$ & $-0.583$ \\
		\hline
		Transition redshift ($z_t$) & 0.79 & 0.73 \\
		\hline
		AIC & 1896.85 & 1908.153 \\
		\hline
		BIC & 1913.254 & 1919.032 \\
		\hline
	\end{tabular}
	\label{tab:mcmc_results}
\end{table}

%%%%%%%%%%%%%%%%%%%%%%%%%%%%%%%%%%%%%%%%%%%%%%%%%%%%%%%%%%%%%%
%%%%%%%%%%%%%%%%%%%%%%%%%%%%%%%% FIG 6 %%%%%%%%%%%%%%%%%%%%%%%%%%%%%%%%%%
\begin{figure}[H]
	\centering
	\includegraphics[scale=0.50]{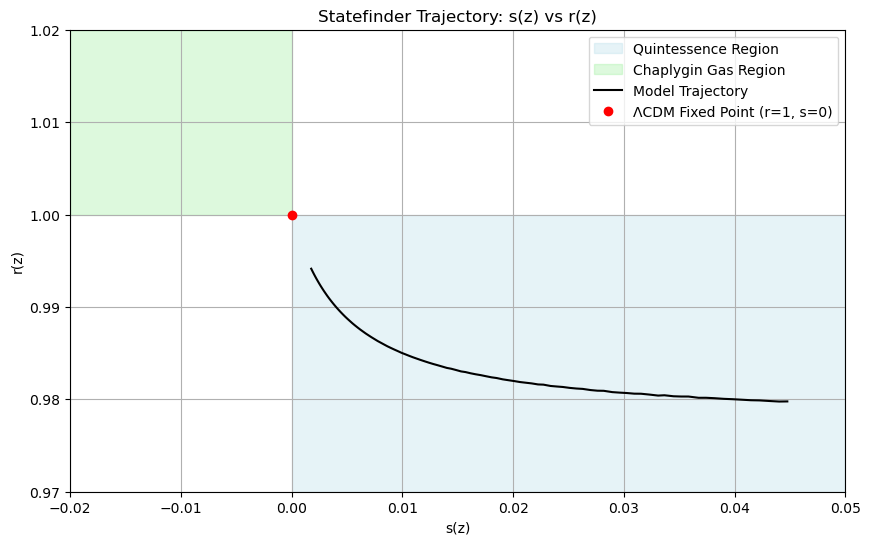}
	\includegraphics[scale=0.50]{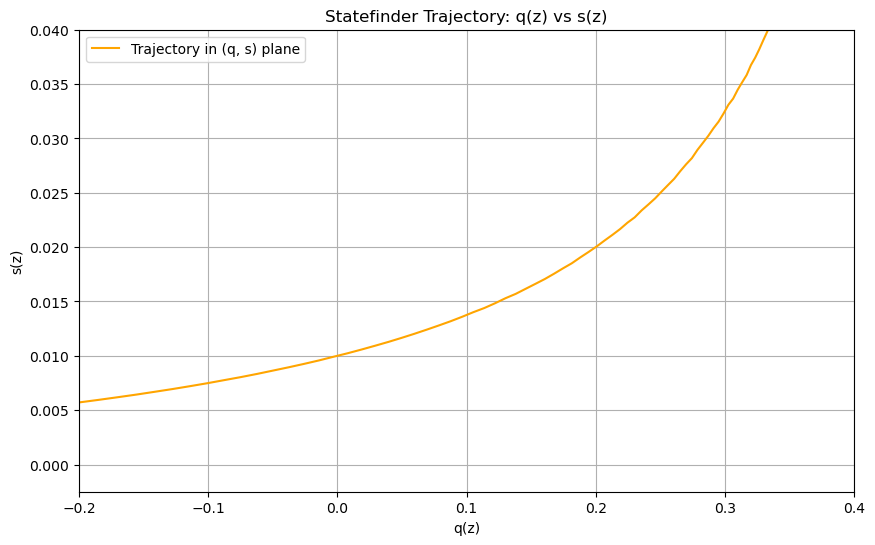}
	\caption{(a) Plot of $r$ vs $s$. (b) Plot of $r$ vs $q$.}
	\label{fig:statefinder}
\end{figure}
%%%%%%%%%%%%%%%%%%%%%%%%%%%%%%%%%%%%%%%%%%%%%%%%%%%%%%%%%%%%
\appendix
\section{Statefinder Diagnostic Analysis}
\label{appendix:statefinder}

The statefinder parameters \((r, s)\) provide a geometrical diagnostic tool to distinguish different dark energy models through higher derivatives of the scale factor \(a(t)\). They are defined as follows \cite{Sahni2003}:\\
$r = \frac{\dddot{a}}{a H^3}, \quad s = \frac{r - 1}{3(q - 1/2)}$,\\
where \(H = \frac{\dot{a}}{a}\) is  Hubble parameter, and \(q = -\frac{\ddot{a}}{a H^2}\) is the deceleration parameter.\\
For the proposed \(f(R, L_m)\) gravity model with the Hubble parameter\\
$H(z) = H_0 \sqrt{(1 - \lambda) + \lambda (1 + z)^{3(1 + w)}}$.\\
These parameters can be expressed in terms of redshift \(z\) and model parameters \(\lambda\) and \(w\). The explicit forms are:\\
$q(z) = -1 + \frac{3}{2} \frac{\lambda (1 + w) (1 + z)^{3(1 + w)}}{(1 - \lambda) + \lambda (1 + z)^{3(1 + w)}}$,\\
$r(z) = 1 + \frac{9}{2} \frac{\lambda (1 + w) (1 + z)^{3(1 + w)}}{(1 - \lambda) + \lambda (1 + z)^{3(1 + w)}} \left[w (1 + w) - \frac{\lambda (1 + w) (1 + z)^{3(1 + w)}}{(1 - \lambda) + \lambda (1 + z)^{3(1 + w)}} \right]$,\\
$s(z) = \frac{r(z) - 1}{3 \left(q(z) - \frac{1}{2}\right)}$.\\
The statefinder trajectory \((s, r)\) for the model closely approaches the \(\Lambda\)CDM fixed point \((0,1)\) at late times, confirming consistency with the standard cosmological paradigm. However, subtle deviations at higher redshifts provide a useful tool to discriminate this model from \(\Lambda\)CDM and other dark energy scenarios.\\
This analysis complements the observational constraints presented in the main text by offering an independent geometric perspective on the cosmic dynamics predicted by the \(f(R, L_m)\) model.

\section{Hubble Table:}
\label{appendix:Hubble Table}
\begin{table}[htbp]
	\centering
	\caption{Hubble Parameter Measurements from Cosmic Chronometers}

		\begin{tabular}{|c|c|c|c|}
			\toprule
			Redshift ($z$) & $H(z)$ [km/s/Mpc] & Error [km/s/Mpc] & Reference \\
			
			0.07 & 69.0 & 19.6 & \cite{Zhang2014} \\
			0.09 & 69.0 & 11.9991 & \cite{Simon2005} \\
			0.12 & 68.6 & 26.2 & \cite{Zhang2014} \\
			0.17 & 83.0 & 8.00037 & \cite{Simon2005} \\
			0.1791 & 74.91 & 3.8069 & \cite{Moresco2012} \\
			0.1993 & 74.96 & 4.9001 & \cite{Moresco2012} \\
			0.2 & 72.9 & 29.6 & \cite{Zhang2014} \\
			0.27 & 77.0 & 13.9986 & \cite{Simon2005} \\
			0.28 & 88.8 & 36.6 & \cite{Zhang2014} \\
			0.3519 & 82.78 & 13.9484 & \cite{Moresco2012} \\
			0.3802 & 83.0 & 13.54 & \cite{Moresco2016} \\
			0.4 & 95.0 & 16.9955 & \cite{Simon2005} \\
			0.4004 & 76.97 & 10.18 & \cite{Moresco2016} \\
			0.4247 & 87.08 & 11.24 & \cite{Moresco2016} \\
			0.4497 & 92.78 & 12.9 & \cite{Moresco2016} \\
			0.4783 & 80.91 & 9.044 & \cite{Moresco2016} \\
			0.47 & 89.0 & 49.6 & \cite{Ratsimbazafy2017} \\
			0.48 & 97.0 & 62.0024 & \cite{Stern2010} \\
			0.5929 & 103.8 & 12.4975 & \cite{Moresco2012} \\
			0.6797 & 91.6 & 7.9619 & \cite{Moresco2012} \\
			0.75 & 98.8 & 33.6 & \cite{Borghi2022} \\
			0.75 & 105.0 & 10.76 & \cite{Jimenez2023} \\
			0.7812 & 104.5 & 12.1951 & \cite{Moresco2012} \\
			0.8 & 113.1 & 31.2 & \cite{Jiao2023} \\
			0.8754 & 125.1 & 16.7009 & \cite{Moresco2012} \\
			0.88 & 90.0 & 39.996 & \cite{Stern2010} \\
			0.9 & 117.0 & 23.0022 & \cite{Simon2005} \\
			1.037 & 153.7 & 19.6736 & \cite{Moresco2012} \\
			1.26 & 135.0 & 65.0 & \cite{Tomasetti2023} \\
			1.3 & 168.0 & 17.0016 & \cite{Simon2005} \\
			1.363 & 160.0 & 33.58 & \cite{Moresco2015} \\
			1.43 & 177.0 & 18.0009 & \cite{Simon2005} \\
			1.53 & 140.0 & 14.0 & \cite{Simon2005} \\
			1.75 & 202.0 & 39.996 & \cite{Simon2005} \\
			1.965 & 186.5 & 50.43 & \cite{Moresco2015} \\
			\hline\hline
		\end{tabular}
	
	\label{tab:hubble_data}
\end{table}

\section*{Declaration of competing interest}
The authors declare that they have no known competing financial interests or personal relationships that could have appeared to influence the work reported in this paper.

	\section*{Acknowledgments}
	The authors (GKG and AP) gratefully acknowledge the facilities and stimulating research environment provided by the Inter-University Centre for Astronomy and Astrophysics (IUCAA), Pune, during their research visits. The author (AP) also expresses gratitude to the University of Zululand, South Africa, for the facilities that were used to complete a part of this work.

\end{document}